# Gate-Controlled P-I-N Junction Switching Device with Graphene Nanoribbon

Shu Nakaharai*, Tomohiko Iijima[1], Shinichi Ogawa[1], Hisao Miyazaki[2], Songlin Li[2],

Kazuhito Tsukagoshi[2], Shintaro Sato, and Naoki Yokoyama

Collaborative Research Team Green Nanoelectronics Center (GNC),

[1]Innovation Center for Advanced Nanodevices (ICAN),

National Institute of Advanced Industrial Science and Technology (AIST),

Tsukuba, Ibaraki 305-8569, Japan

[2]International Center for Materials Nanoarchitectonics (MANA),

National Institute for Materials Science (NIMS), Tsukuba, Ibaraki 305-0044, Japan

*E-mail address: shu-nakaharai@aist.go.jp

Abstract :

A graphene P-I-N junction switching device with a nanoribbon is proposed, which was aimed at finding an optimized operation scheme for graphene transistors. The device has two bulk graphene regions where the carrier type is electrostatically controlled by a top gate, and these two regions are separated by a nanoribbon that works as an insulator, resulting in a junction configuration of (P or N)-I-(P or N). It is demonstrated that the drain current modulation strongly depends on the junction configuration, while the nanoribbon is not directly top-gated, and that the device with a P-I-N or N-I-P junction can exhibit better switching properties.

Graphene[1,2] has attracted considerable attention for its various unique features, such as its extremely high mobility up to 200,000 $cm^2/(Vs)$[3,4], ambipolar carrier properties that can be controlled electrostatically[5], and atomically thin body channels that enable transistors to be aggressively scaled[6]. These advantages should enable future superhigh-speed or ultralow-power-consumption LSIs to be fabricated, while the lack of band gap is still the most serious bottleneck. The most promising techniques for overcoming the band gap issue are graphene nanoribbons (GNRs)[7–11] or bilayer graphene under a high electric field perpendicular to the sheet[12]. However, the magnitude of the band gap is not large enough for sufficient cut-off properties. Therefore, an optimized operation scheme for graphene field-effect transistors is urgently required, which would enable a high on-off ratio.

This paper proposes the concept of a graphene device aimed at inducing superior on-off properties based on the GNR technique, and demonstrates its basic operation in an experiment. The proposed device has two bulk graphene regions under local top gates, and between them lies a GNR, as illustrated in Fig. 1(a). The carrier types are independently controlled in the two top-gated bulk regions by biasing into the P or N type. The GNR behaves as an insulator when the Fermi level is in an energy gap[10,11]. Thus, the device has the structure of a (P or N)-I-(P or N) junction whose junction type can be electrostatically controlled.

The basic operations of the proposed device are outlined in Figs. 1(b)–1(d). The Fermi level is higher in the on state [Fig. 1(c)] than the top of the energy gap. The off state can be obtained by lifting the band of the bulk graphene region under the top gate (TG1) by flipping the top-gate bias from positive to negative [Fig. 1 (d)]. Electrons in the source region will experience a tall barrier height (indicated by the vertical arrow) and a long barrier length (indicated by the horizontal arrow) for transport to the drain as shown in Fig. 1(d). This off state can reduce the off-leak current drastically compared with the conventional graphene device in Fig. 1(g). Electron (or hole) transport in this conventional case is blocked by a barrier at the edge of the gated GNR, and off-leak current is easily induced by thermally-activated carriers going over the barrier. Here, at most half of the gap energy contributes to the off state, because both electrons and holes can be transported through the nanoribbon [Fig. 1 (g)]. However, the barrier height in the proposed device can be much higher than the band gap. For example, the barrier height should be ~340 meV at $|V_{TG1}|$ = 1 V in a 5-nm-thick $SiO_2$ gate dielectric layer. In this paper, we demonstrate the strong gating of the GNR by raising the band of the bulk graphene, which is proximate to the GNR. We also demonstrated that the current through the device was modulated according to the junction configurations of P-I-N, P-I-P, N-I-P, and N-I-N. These experimental results are good demonstrations of the possibility of achieving graphene P-I-N junction transistors.

The fabrication of the graphene devices for the demonstration was carried out as follows. Single-layer graphene flakes were mechanically exfoliated from a highly oriented pyrolytic graphite (HOPG) crystal with adhesive tape[1)], and deposited on a silicon wafer with a 285-nm-thick surface thermal oxide layer. The location of the graphene flakes was identified by sight with an optical microscope[13)], and then source/drain contacts were attached using electron beam lithography with PMMA resists, thermal evaporation of Ti/Au (= 5 nm/30 nm), and a lift-off process. The contact resistance was evaluated to be sufficiently small (~1 kΩ) by two-terminal resistance measurements with back gate bias sweeping, and every device had a distinct resistance peak at only a few volts from the zero of back-gate bias.

The top-gate stacks, which consisted of a 5-nm-thick $SiO_2$ layer and a 20-nm-thick Al layer, were also fabricated by electron beam lithography and the lift-off process [Fig. 2(a) inset]. The current modulation by one of the top-gates' bias was found to be as small as 10% of the whole drain current, indicating that the resistance of only a small fraction of the graphene region was changed by the top gate. The leak currents of the top gates were lower than the detection limit of 1 pA in both top gates. The two top gates form a point-contact-like structure with a gap length of about 50 nm as shown in the inset of Fig. 2(b). The structure was fabricated by lithography of a pair of triangle patterns contacting to each other at an apex, and the gate gap was opened near the contact point of the triangles where the pattern size was smaller than the resolution of lithography. The graphene was covered by $SiO_2$ etching masks except for the region between the top gates, and then the uncovered graphene was etched off using oxygen plasma at 60 W for 1 min. After etching, the graphene remained in the gate gap, forming a GNR with a width of about 40 nm as shown in the inset of Fig. 2(b), although this area was not actually covered by any protective layers. One possible reason for the graphene surviving the etching process is that a small amount of redeposited residue covered the graphene surface in the etching process and protected the graphene from being etched. Here, the direction of the edge was not controlled specially, and it is supposed that the edge of the GNR has high density of disorders.

Transport measurements were carried out in a cryogenic probe station under a high vacuum on the order of $10^{-5}$ Pa. Figure 2(b) shows on-off operation by sweeping the bias of TG1 with fixed biases for the drain ($V_d$ = 1 mV), TG2 ($V_{TG2}$ = 4 V), and back gate ($V_{BG}$ = 0 V). As seen in the figure, the on-off ratio is about one order of magnitude. Here, most of the change in the resistance from ~100 kΩ to ~1 MΩ is attributed to the change in resistance associated with the GNR, while the GNR itself is not directly gated. This is because the resistance of the bulk graphene under TG1 is expected to be as small as on the order of 1 kΩ as detected and shown in Fig. 2(a). This proves that the band configuration for the GNR is controlled by the gate bias of the bulk graphene region under TG1, which are proximate to the GNR. This proximate gating effect on the GNR is good evidence for the possibility of the device concept that we proposed. Here, spike-like peaks in the $V_{TG1}$-$I_d$ curve in Fig. 2 (b) were reproduced by multiple measurements, suggesting that these signals reflect the resonant conduction through localizing sites which are proper to each GNR[10)].

Another demonstration of our device is the dependence of drain current on junction types, such as P-I-N and P-I-P. Figure 3(a) shows the drain currents in all types of junction configuration in another device by sweeping the TG1 bias, $V_{TG1}$, at $V_{TG2} = \pm 4$ V. For the P-I-P (or N-I-N) configuration in Fig. 3(b) [Fig. 3 (d)], the Fermi level is higher (lower) than the top (bottom) of the energy gap induced in the GNR, meaning that there is no barrier for carrier transport through the junctions. For N-I-P and P-I-N, on the other hand, the Fermi level crosses the energy gap in the GNR part [Fig. 3(c) and 3(e)], resulting in lower drain currents than those for P-I-P and N-I-N as expected. Obviously, the TG1 bias modulation of the drain currents depends on the polarity of the TG2 bias, which is separated from the TG1 region by the GNR. This strongly proves that the current through the P-I-N junction is dependent on the junction configuration, i.e., the basic principle underlying the operation of our device.

The minimum drain current in the measurement of Fig. 3 (a) increased slightly as the temperature increased (data not shown). From these data, the activation energy was estimated to be 10 meV, which is usually referred to as the band gap. This gap energy corresponds to a reasonable effective ribbon width of about 50 nm[10].

The P-I-N junction configuration in the high-drain-bias regime has another effect on carrier transport according to its direction, which is negligible at low drain biases. Figures 4(a) and 4(b) show the absolute values for drain currents at $V_d$ = ± 10 mV and $V_d$ = ± 100 mV, respectively. The $|I_d|$-$V_{TG1}$ curve in Fig. 4(a) is identical regardless of the current direction. For a high drain bias of $V_d$ = ± 100 mV [Fig. 4(b)], however, the P-I-N or N-I-P configuration was found to split into two current levels according to the current direction. As can easily be seen in Fig. 4(b), the electrons flowing in the order of P→I→N are always higher than those for N→I→P. This effect can be explained in terms of the difference in the barrier lengths in the GNR indicated by the horizontal arrows in Figs. 4(c) to 4(f). The steep slope of the GNR band for the configurations in Figs. 4(c) and 4(e) shortens the barrier length more than that in Figs. 4(d) and 4(f), resulting in a larger leak current through the potential barrier.

All these data are consistent with what were expected from the dependence of drain current on the P-I-N junction configuration. Therefore, the basic concept behind the operation of our device is considered to be feasible. Although the low on/off ratio of one order of magnitude in the present results is insufficient for logic devices, narrower GNRs than that of the present device can result in much better on-off properties due to the larger energy gap[10,11]. In that sense, it is fair to expect the proposed P-I-N junction switching device in which the off state will be much more effective than the conventional device structure if compared at the same energy gap strength. In addition, the proposed device structure also contributes to larger on current than the conventional one by shortening the length of the low-mobility GNR.

In summary, we proposed a graphene nanoribbon switching device with a P-I-N

junction and demonstrated its operation. We proved that the drain current was modulated by controlling the junction type as was expected from the whole band configuration. These results should contribute to further developments in graphene device technology by offering the possibility of a GNR channel device structure with a high on-off ratio.

Acknowledgement: This research is granted by JSPS through FIRST Program initiated by CSTP.

Captions:

**Fig. 1** (a) Schematic of proposed graphene nanoribbon device. (b) Top view of device. (c) Band diagrams for proposed device in the on state. (d) Off state is obtained by flipping the polarity of either $V_{TG1}$ or $V_{TG2}$. Here, small horizontal arrows represent the barrier length and vertical arrows represent the barrier height for carrier transport. Conventional GNR transistor and its on-off operation are shown in (e), (f), and (g). In the off state (g), thermally activated electrons and holes easily flow over the potential barrier, resulting in a large off leakage.

**Fig. 2** (a) Current modulation by two top gates at room temperature before a nanoribbon is etched. Inset is an optical micrograph of the device. Dotted lines indicate the boundary of a single-layer graphene. The width of the graphene is 1.1 – 1.4 μm. (b) $V_{TG1}$ dependence of drain current in same device after nanoribbon is etched with $V_{TG2}$ = +4 V and $V_d$ = +1 mV at $T$ = 45 K. Inset is a helium ion micrograph of the nanoribbon part of the device. The bright region in the gap of the split gate is a graphene nanoribbon.

**Fig. 3** (a) Top gate modulation of drain current in P-I-N junction device at $V_d$ = +1 mV and $T$ = 45 K. The $I_d$-$V_{TG1}$ curve at fixed $V_{TG2}$ = +4 V is in red, and corresponding junction configurations are illustrated in (c) and (d). Similarly, the curve at fixed $V_{TG2}$ = −4 V is in blue (broken) and illustrated in (b) and (e).

**Fig. 4** $|I_d|$-$V_{TG1}$ curves under low (a) and high (b) drain bias conditions. The $|I_d|$-$V_{TG1}$ curve behaves similarly regardless of current direction in (a), while it does not in (b). Band configurations for four off states are illustrated in (c)–(f). The difference in current strength according to direction is attributed to the difference in barrier length indicated by horizontal arrows.

Fig. 1

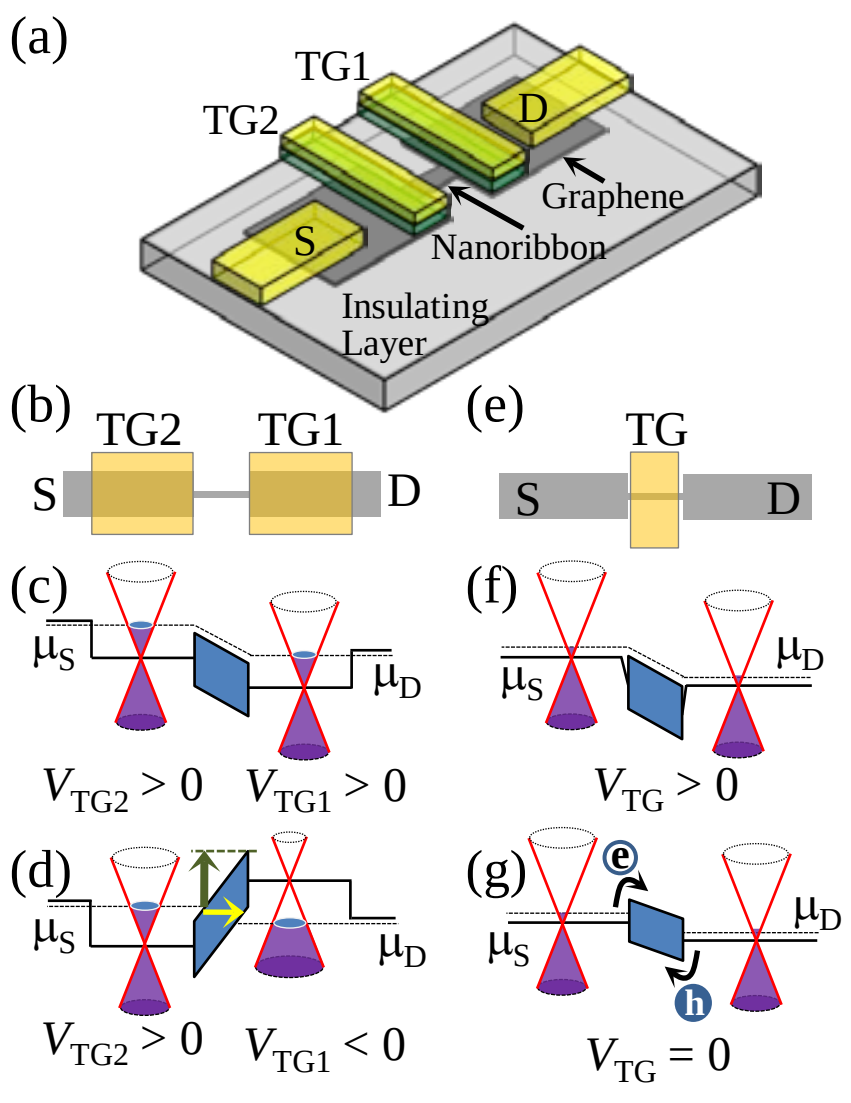

Fig. 2

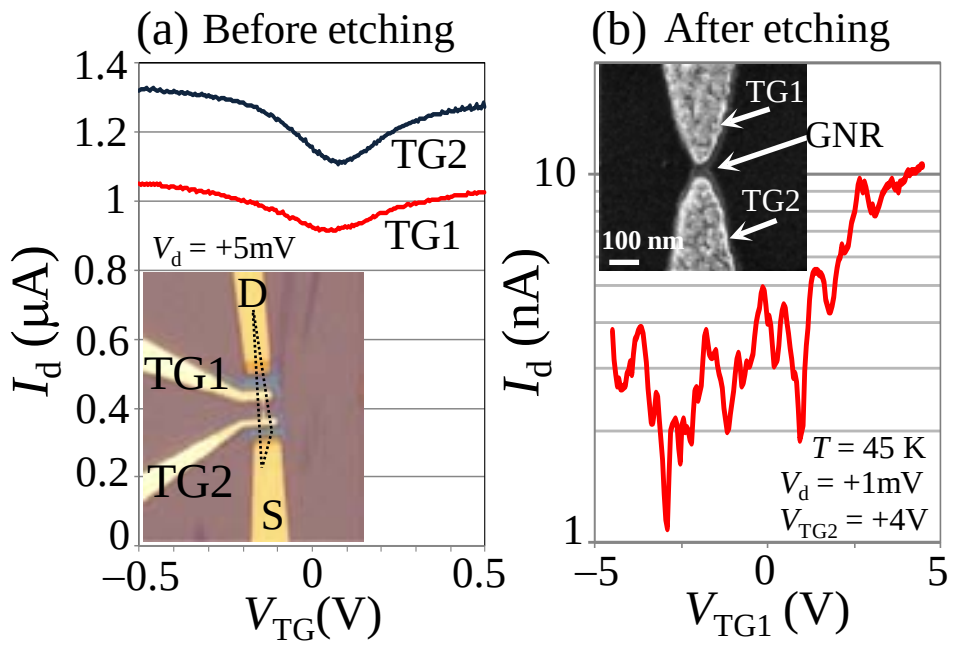

Fig. 3

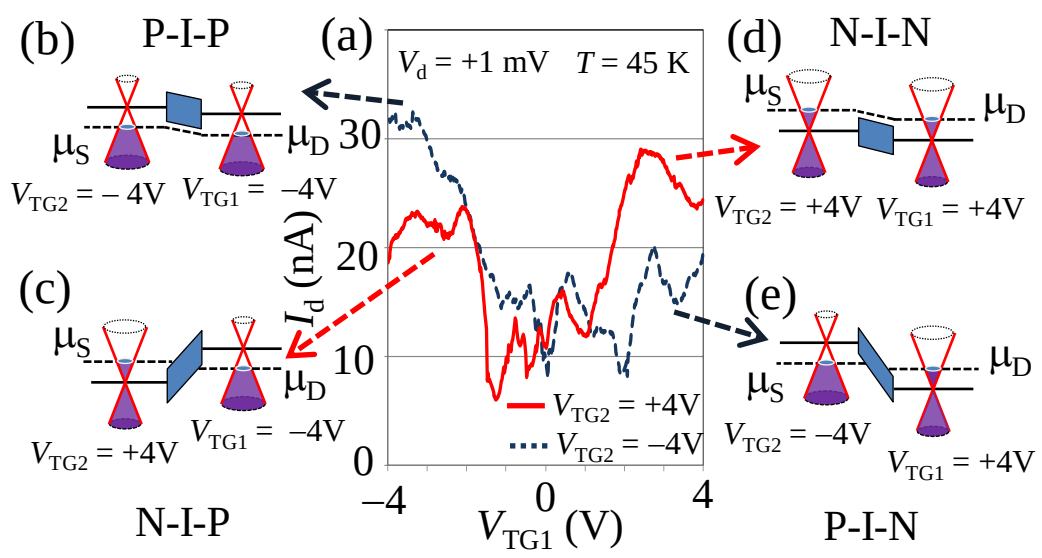

(b)
P-I-P
(a)
$V_d$ = +1 mV $T$ = 45 K
(d)
N-I-N
$\mu_S$
$\mu_D$
$V_{TG2}$ = – 4V $V_{TG1}$ = –4V
$V_{TG2}$ = +4V $V_{TG1}$ = +4V
(c)
(e)
$I_d$ (nA)
30
20
10
0
–4
0
4
$V_{TG1}$ (V)
$V_{TG2}$ = +4V
$V_{TG2}$ = –4V
$V_{TG2}$ = +4V $V_{TG1}$ = –4V
$V_{TG2}$ = –4V $V_{TG1}$ = +4V
N-I-P
P-I-N

Fig. 4

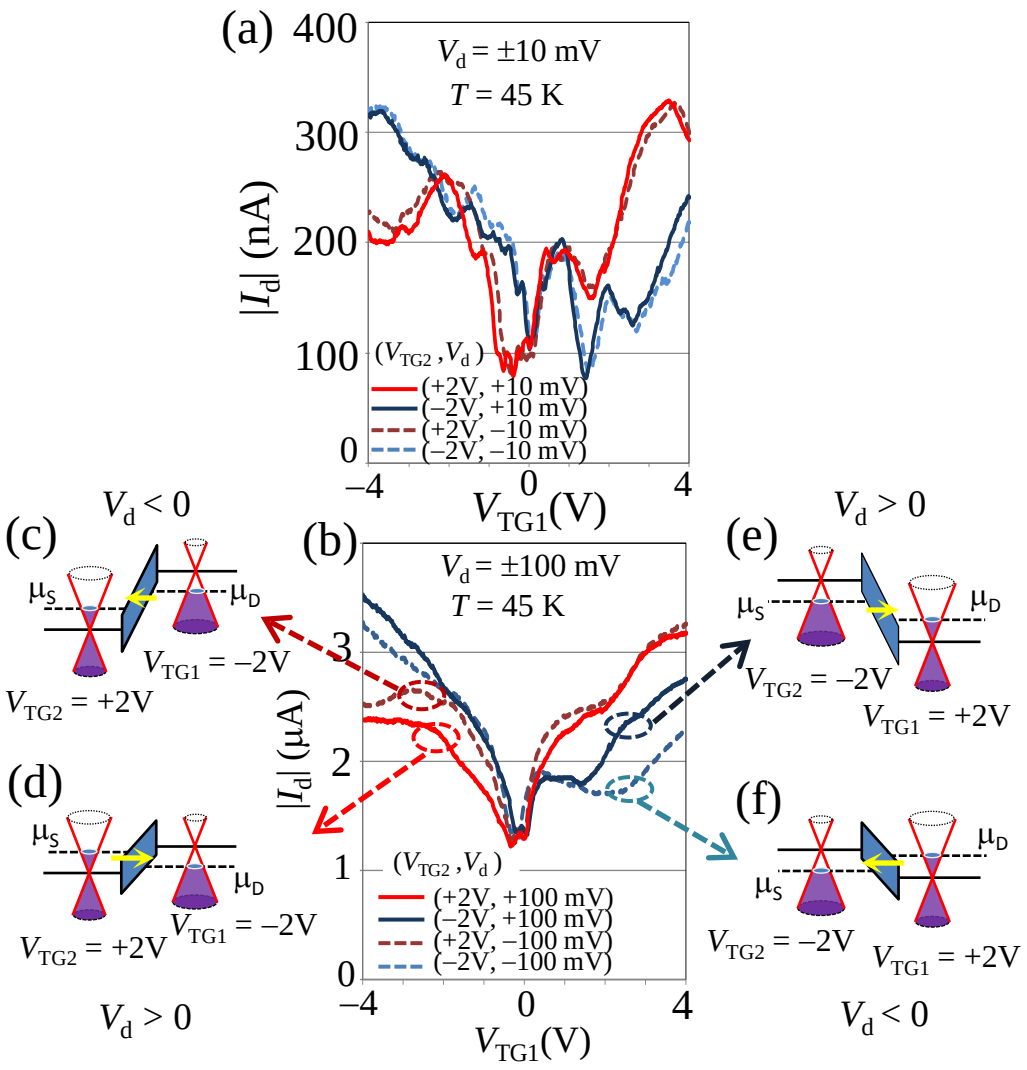